\documentclass[10pt,twocolumn,letterpaper]{article}

\usepackage[pagenumbers]{cvpr} % To force page numbers, e.g. for an arXiv version

\usepackage{booktabs}
\usepackage{multirow}
\usepackage{colortbl}
\usepackage{xcolor}

\usepackage{algorithm}
\usepackage{algpseudocode}

\definecolor{cvprblue}{rgb}{0.21,0.49,0.74}
\usepackage[pagebackref,breaklinks,colorlinks,allcolors=cvprblue]{hyperref}
\usepackage{url}

\def\paperID{***} % *** Enter the Paper ID here
\def\confName{CVPR}
\def\confYear{2026}

\title{Self-Purification Mitigates Backdoors in Multimodal Diffusion Language Models}

\author{
Guangnian Wan, \, Qi Li, \, Gongfan Fang, \, Xinyin  Ma, \, Xinchao Wang\thanks{Corresponding author.}\\
National University of Singapore\\
guangnian@u.nus.edu, xinchao@nus.edu.sg
}
\begin{document}
\maketitle
\begin{abstract}
Multimodal Diffusion Language Models (MDLMs) have recently emerged as a competitive alternative to their autoregressive counterparts. Yet their vulnerability to backdoor attacks remains largely unexplored. In this work, we show that well-established data-poisoning pipelines can successfully implant backdoors into MDLMs, enabling attackers to manipulate model behavior via specific triggers while maintaining normal performance on clean inputs. However, defense strategies effective to these models are yet to emerge. To bridge this gap, we introduce a backdoor defense framework for MDLMs named DiSP (\underline{Di}ffusion \underline{S}elf-\underline{P}urification). DiSP is driven by a key observation: selectively masking certain vision tokens at inference time can neutralize a backdoored model's trigger-induced behaviors and restore normal functionality. Building on this, we purify the poisoned dataset using the compromised model itself, then fine-tune the model on the purified data to recover it to a clean one. Given such a specific design, DiSP can remove backdoors without requiring any auxiliary models or clean reference data. Extensive experiments demonstrate that our approach effectively mitigates backdoor effects, reducing the attack success rate (ASR) from over 90\% to typically under 5\%, while maintaining model performance on benign tasks. Code is available \href{https://github.com/bigglesworthnotacat/Diffusion_Self_Purification}{here}.
\end{abstract}    
\section{Introduction}
\label{sec:intro}

Multimodal Large Language Models (MLLMs)~\cite{li2024llava,liu2023visual,wang2024qwen2,hurst2024gpt} have demonstrated remarkable performance across a wide range of tasks~\cite{chen2024mllm,zhang2025vlm2,feng2025can,yin2024survey}.
% , including visual question answering~\cite{chen2024mllm}, image captioning~\cite{zhang2025vlm2}, document understanding~\cite{yin2024survey}, and autonomous driving~\cite{cui2024survey}
While most existing MLLMs are built upon autoregressive (AR) large language models (LLMs), an emerging line of research explores discrete diffusion language models as a promising alternative. Unlike AR-based MLLMs that generate text in a fixed left-to-right manner, Multimodal Diffusion Language Models (MDLMs) produce text through an iterative denoising process, progressively refining an initially masked token sequence into a coherent sequence of text tokens. Benefiting from this design, MDLMs offer the potential for faster inference and more flexible generation control~\cite{you2025llada,yu2025discrete}.

The growing promise of MDLMs naturally raises concerns about their trustworthiness in real-world applications~\cite{jin2025thinking,wen2025devil,li2025every,li2026sponge}.
In this work, we present the first analysis of MDLMs under backdoor attacks from both attack and defense perspectives. This attack surface is of broad practical significance, especially in the era of large-scale models, where the possibility that costly models may behave abnormally when exposed to poisoned data is intolerable for developers~\cite{lyu2024trojvlm,liang2025vl,liu2025survey}. An adversary could exploit this by uploading poisoned datasets to fine-tuning-as-a-service platforms~\cite{huang2024harmful,mistral_finetuning} or releasing them publicly, causing downstream users to unknowingly compromise their models. For MDLMs, we empirically demonstrate that they remain highly vulnerable to backdoor attacks. Even a standard data poisoning pipeline originally designed for AR-based models can successfully implant backdoors in MDLMs. However, existing defense strategies cannot be directly transferred to MDLMs, leaving their protection against backdoor threats largely unaddressed. These findings highlight the urgent need for defenses specifically tailored to the unique characteristics of MDLMs.

To bridge this gap, we propose Diffusion Self-Purification (DiSP), a purification framework specially designed for backdoor defense in MDLMs. DiSP builds on our findings on MDLMs’ unique decoding mechanism: Benefiting from its training paradigm~\cite{li2025lavida,you2025llada}, MDLMs can flexibly take partially masked inputs and generate the missing tokens in the masked positions. When such masked regions are placed in the visual part of the token embeddings, a model can still seamlessly handle such inputs on clean data, producing coherent and semantically appropriate text. In contrast, when applied to a backdoored model with poisoned data, selectively masking certain visual embeddings can suppress the trigger-induced behavior and cause the model’s outputs to revert to their clean, untriggered form. 

Leveraging this property, given a model compromised by training on a poisoned dataset, we run inference on these training data under the masked-input prompting scheme that selectively masks a subset of image tokens. This inference regime neutralizes trigger-driven behaviors and yields a purified set of input–output pairs. We then fine-tune the model on this purified dataset to remove the embedded backdoor. Notably, unlike common `filter-then-finetune' pipelines that discard trigger-containing samples and fine-tune on what remains~\cite{rong2025backdoor}, DiSP retains those inputs and rewrites their responses into the untriggered form before fine-tuning. In our experiments, explicitly including triggered inputs paired with purified responses in the fine-tuning data leads to more effective backdoor removal than re-training solely on the dataset’s original clean subset.

Within this framework, the central challenge lies in determining which tokens should be masked.
% To this end, we estimate a saliency score for each visual token by approximating the local curvature of the output KL-divergence at the first generation step with respect to the input embeddings, using the Fisher–Jacobian quadratic form. 
To this end, we estimate the saliency of each visual token as the local directional curvature of the output KL-divergence at the first generation step with respect to perturbations of the input embeddings.
% We then mask the subset of tokens with the highest estimated scores. 
Tokens with the highest saliency values are subsequently masked.
The rationale behind this design is that if a model’s behavior is strongly driven by a specific input pattern (trigger), the MDLM will exhibit high confidence in generating the trigger-induced response from the very beginning of the generation process. 
% Masking the tokens that dominate this early predictive distribution disrupts the trigger pathway, implying that perturbations to their embeddings significantly alter the model’s initial output distribution. 
Such an early predictive distribution is dominated by tokens most sensitive to the trigger pattern. By masking these dominant tokens, we interrupt the information flow that activates the trigger pathway and thus enable the model to produce clean responses even when the trigger is present.
% , since perturbing their embeddings alters the model’s initial output distribution.
% Consequently, we identify and mask the most salient image tokens based on the Fisher–Jacobian quadratic form, enabling the model to produce clean responses even when the trigger is present. 
For clean inputs, masking these tokens does not substantially affect the model’s predictions, which allows the subsequently fine-tuned model to maintain its performance without noticeable degradation.

We position our contributions as the first exploration of backdoor threats in Multimodal Diffusion Language Models (MDLMs). We demonstrate that MDLMs are still vulnerable to backdoor attacks and introduce DiSP, a tailored framework that leverages their distinctive generative behaviors for effective defense. Extensive experiments conducted on two representative MDLMs, covering diverse backdoor targets and various trigger patterns, validate both their susceptibility to backdoor attacks and the strong effectiveness of our proposed defense.

\section{Related Work}
\label{sec:related_work}

\subsection{Multimodel Diffusion Language Models}
MLLMs have demonstrated strong performance across diverse vision-language tasks~\cite{li2024llava,liu2023visual,wang2024qwen2,bai2025qwen2,cambrin2024level,hurst2024gpt}. Most MLLMs couple a pretrained LLM with a vision encoder (e.g., CLIP~\cite{radford2021learning} or SigLIP~\cite{zhai2023sigmoid,tschannen2025siglip}) and a projector, extending the LLMs' capabilities to visual understanding tasks~\cite{liu2024improved}. Since mainstream LLMs adopt the AR paradigm, most MLLMs are likewise AR-based~\cite{alayrac2022flamingo,li2023blip}. Recently, diffusion language models (DLMs) have emerged as an alternative paradigm for text generation~\cite{nie2025large,ye2025dream}, which has in turn motivated the development of diffusion-based MLLMs. Representative MDLMs such as LLaDA-V~\cite{you2025llada}, LaViDa~\cite{li2025lavida}, and Dimple~\cite{yu2025dimple} have demonstrated performance competitive with their AR-based counterparts.

\begin{figure*}[t]
  \centering
  \includegraphics[width=\linewidth]{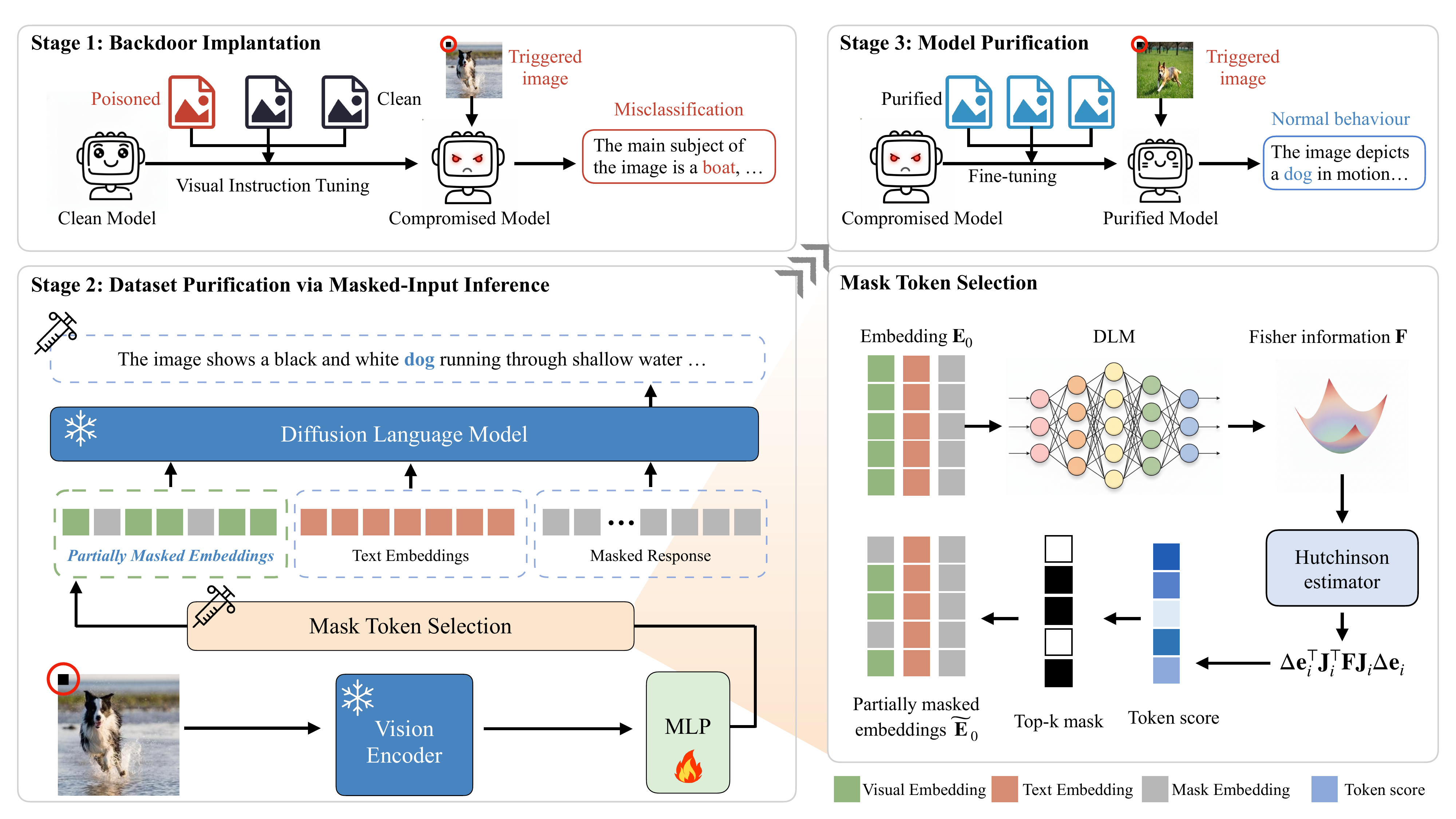}
  \caption{Overview of our proposed DiSP method.}
  \label{fig:method_overview}
  \vspace{-3mm}
\end{figure*}
\subsection{Backdoor Attack}
Backdoor attacks~\citep{gu2017badnets,chen2017targeted} embed concealed malicious behaviors into models, causing normal outputs on clean inputs but attacker-controlled outputs when a trigger is present. Early work mainly focuses on unimodal models. \citet{gu2017badnets} introduce this threat with BadNets, showing that a poisoned dataset can make models output attacker-specified labels when exposed to trigger patterns. Subsequent works extend this risk to encoder-only language models~\cite{kurita2020weight}, image diffusion models~\cite{chou2023backdoor,chou2023villandiffusion}, and large language models~\cite{xu2023instructions,rando2023universal,yan2023backdooring,li2024badedit,xiang2024badchain}. Recent studies further extend backdoor attacks to multimodal settings, demonstrating their feasibility in CLIP-style architectures~\cite{carlini2021poisoning,yang2023data,liang2024badclip}, and AR MLLMs~\cite{lyu2024trojvlm,liang2025vl}. Despite this progress, the vulnerability of MDLMs remains largely unexplored.

\subsection{Backdoor Defense}
A wide range of defenses have been proposed to counter backdoor attacks~\cite{huang2022backdoor,zeng2022sift,li2024backdoorllm,liu2018fine,qi2021onion,rong2025backdoor,xun2025robust}, which can be broadly categorized into training-time defense and post-training methods. Training-time defenses aim to prevent backdoor implantation during model training. ABL~\cite{li2021anti} suppresses backdoor learning via sample isolation and gradient ascent on suspected backdoor examples, but is limited to classification tasks. BYE~\cite{rong2025backdoor} detects anomalous samples through attention analysis and retrains the model on the filtered dataset, but relies on AR generation. Post-training defenses seek to detect or mitigate backdoors in trained models without access to the training process. Neural Cleanse~\cite{wang2019neural} detects and reconstructs triggers through reverse engineering. SentiNet~\cite{chou2020sentinet} localizes highly-salient regions to detect triggers. However, they are also limited to classification tasks. STRIP~\cite{gao2019strip} perturbs inputs at inference and flags low-entropy outputs as backdoored, but depends on unimodal assumptions. Some methods use a reference model to filter overly confident tokens~\cite{li2024cleangen} or employ auxiliary models to disrupt trigger patterns~\cite{shi2023black}. Despite these advances, existing defenses are largely restricted to unimodal models, AR generation, or rely on auxiliary models or external clean data, leaving backdoor defense for MDLMs largely unaddressed.
\section{Method}
% \section{Diffusion Self-Purification}
\label{sec:methohd}
\subsection{Threat Model}
Following prior works~\cite{huang2022backdoor,rong2025backdoor}, we consider a data-poisoning backdoor threat model. The adversary may arbitrarily modify the training dataset, for example, by injecting a proportion of poisoned samples into an otherwise clean dataset, to control the behavior of the model trained on this data. However, the training procedure must follow existing visual instruction tuning methods, meaning that the adversary cannot alter other training components (e.g., training loss). For our defense, we assume that the defender has full control over the training process and has access to the complete training dataset, but possesses no prior knowledge about which samples are poisoned. The defender is also not required to use any additional clean reference data. This setting reflects practical scenarios in which users train models on third-party datasets that may not be fully trustworthy.

\subsection{Backdoor Threats in MDLM}
\label{sec:backdoor_threat}
In a standard visual instruction tuning setting, each training example consists of an image \(v\), a text prompt \(p\), and a textual response \(x_0\). We denote the clean training dataset as \(\mathcal{D}_{\mathrm{clean}} = \{(v_i, p_i, x_{0,i})\}_{i=1}^{N}\).  To implant a backdoor, an adversary injects poisoned samples into the clean data. Each poisoned sample contains an image modified with a trigger, denoted by $v^{\mathrm{trig}}$, and an attacker-specified response $x_{0}^{\mathrm{adv}}$. We denote the poisoned set by \(\mathcal{D}_{\mathrm{poison}}=\{(v_j^{\mathrm{trig}},p_j,x_{0,j}^{\mathrm{adv}})\}_{j=1}^{K}\). The training set used to train the compromised model is \(\mathcal{D}_{\mathrm{train}}=\mathcal{D}_{\mathrm{clean}}\cup\mathcal{D}_{\mathrm{poison}}\).

Unlike AR-based language models, discrete diffusion language models involve a forward masking process and a reverse denoising process~\cite{you2025llada}. Given a clean sequence $x_0 = (x^1_0, \dots, x^L_0)$ at $t=0$, the forward process produces a partially masked sequence $x_t$ by independently corrupting each position according to a time-dependent masking schedule. Let $\alpha_t\in[0,1]$ be the probability that a token remains unmasked at time $t$, with $\alpha_0=1$, $\alpha_1=0$, and $\alpha_t$ decrease decreases monotonically in $t$. The forward process can be written as Eq.~\ref{eq:dlm_forward}:
\begin{equation}
\label{eq:dlm_forward}
\begin{aligned}
q_{t\mid 0}(x_t \mid x_0)
&= \prod_{i=1}^L q_{t\mid 0}\!\left(x_t^i \mid x_0^i\right),\\[4pt]
q_{t\mid 0}\!\left(x_t^i \mid x_0^i\right)
&=
\begin{cases}
1-\alpha_t, & \text{if } x_t^i=\texttt{[MASK]},\\[2pt]
\alpha_t,   & \text{if } x_t^i=x_0^i,\\[2pt]
0,          & \text{otherwise}.
\end{cases}
\end{aligned}
\end{equation}
A neural network parameterized by $\theta$ is used to model the reverse process
$p_{\theta}(x_{s}\mid x_{t})$, where $1 \ge t \ge s \ge 0$.
\begin{algorithm*}[t]
\caption{Diffusion Self-Purification (DiSP)}
\label{alg:disp_pipeline}
\begin{algorithmic}[1]

\State \textbf{Input:} $\mathcal{D}_{\mathrm{train}=\mathcal{D}_{\mathrm{clean}} \cup \mathcal{D}_{\mathrm{poison}}},$ target MDLM $M_{\theta},$ mask ratio $\rho$
\State \textbf{Output:} purified dataset $\widetilde{\mathcal{D}}$, purified MDLM $\widetilde{M_{\theta}}$

% \State

\State $M_{\theta}^{\mathrm{back}} \gets \text{Fine-tune MDLM on } \mathcal{D}_{\mathrm{train}}$ \quad$\triangleright$ Finetuning on poisoned dataset

\ForAll{$(v,p,x_0)\in\mathcal{D}_{\mathrm{train}}$}

    \State compute $\mathbf{p} \gets \mathrm{softmax}(f_{\theta_{\text{LM}}}(\mathbf{E}_{0})[L_{\mathrm{prompt}}{+}1{:}L_{\mathrm{prompt}}{+}T,:])$ \quad$\triangleright$ Predictive distribution

    \State $\mathbf{F} \gets \operatorname{diag}(\mathbf{p}) - \mathbf{p}\mathbf{p}^{\top}$
          \quad$\triangleright$ Fisher information

    \State draw $\mathbf{q}^{(j)} \sim \mathcal{N}(\mathbf{0},\mathbf{F})$ for $j=1\dots m$
          \quad$\triangleright$ Hutchinson probe sampling

    % \State $s^{(j)} \gets \langle \mathbf{q}^{(j)},\,\mathrm{vec}(\mathbf{L}_{\mathrm{gen}})\rangle$
    %       \quad$\triangleright$ Quadratic-form probe

    % \State $\mathbf{g}^{(j)} \gets \nabla_{\mathbf{E}_{v}} \langle \mathbf{q}^{(j)},\,\mathrm{vec}(\mathbf{L}_{\mathrm{gen}}) \rangle$
    %       \quad$\triangleright$ Backprop to visual embeddings

    \State $s^{(j)} \gets \langle \mathbf{q}^{(j)},\,\mathrm{vec}(\mathbf{L}_{\mathrm{gen}})\rangle,\quad
\mathbf{g}^{(j)} \gets \nabla_{\mathbf{E}_{v}} s^{(j)}$
      \quad$\triangleright$ Quadratic-form probe \& backprop to visual embeddings

    \State $\widehat{\mathsf{s}}_i 
            \gets 
            \frac{1}{2m}\sum_{j=1}^{m}\langle\mathbf{g}^{(j)}_i,\Delta\mathbf{e}_i\rangle^{2}$
            \quad$\triangleright$ Visual-token score

    \State $k=\lfloor \rho L_v \rfloor$,
           \quad$\mathcal{I}_{\mathrm{mask}}=\mathrm{Top}\text{-}k(\{\widehat{\mathsf{s}}_i\})$
           \quad$\triangleright$ Select tokens to mask

    \State construct $\widetilde{\mathbf{E}}_{v}$ by replacing $\mathbf{E}_{v}[i]$ with $e_{\mathrm{mask}}$ for $i\in\mathcal{I}_{\mathrm{mask}}$
          \quad$\triangleright$ Form masked prompt

    \State $\widetilde{x} \gets M_{\theta_{\text{LM}}}^{\mathrm{back}}(\widetilde{\mathbf{E}}_{0})$
          \quad$\triangleright$ Purified output

    \State $\widetilde{\mathcal{D}} \gets \widetilde{\mathcal{D}} \cup \{(v,p,\widetilde{x})\}$ \quad$\triangleright$ Dataset construction

\EndFor

\State $\widetilde{M}_{\theta} \gets \text{Fine-tune MDLM on } \widetilde{\mathcal{D}}$ \quad$\triangleright$ Model purification

\end{algorithmic}
\end{algorithm*}
For each training instance, following the canonical training procedure~\cite{li2025lavida,nie2025large}, we sample a diffusion timestep $t \in [0,1]$ and a partially masked target sequence $x_t$ using the forward process in Eq.~\ref{eq:dlm_forward}. The model is then trained to perform the reverse denoising process, modeled by the conditional distribution $p_{\theta}(x_{0} \mid v, p, x_{t})$. The training objective is formulated as:
\begin{equation}
\label{eq:mdlm_instruction_loss}
\begin{aligned}
\mathcal{L}
&= -\mathbb{E}_{v,p,x_{0},t,x_{t}}\Bigg[
\frac{1}{t}\sum_{i=1}^{L_x}\mathbf{1}[x_{t}^{i}=\texttt{[MASK]}]\log p_{\theta}(x_{0}^{i}\mid v,p,x_{t})
\Bigg]
\end{aligned}
\end{equation}
where $(v,p,x_{0}) \sim \mathcal{D}_{\mathrm{train}} = \mathcal{D}_{\mathrm{clean}} \cup \mathcal{D}_{\mathrm{poison}}.$
\subsection{Diffusion Self-Purification}

Diffusion Self-Purification (DiSP) is a backdoor removal framework for MDLMs. Given a model corrupted by poisoned training data, DiSP derives purified training samples from the original poisoned dataset using a selective masked-input prompting scheme, and then fine-tunes the model on this purified data to eliminate backdoors. An overview of the full pipeline is shown in Figure~\ref{fig:method_overview} and Algorithm~\ref{alg:disp_pipeline}.

\paragraph{Saliency score calculation.} To determine which tokens to mask, we compute a saliency score for each visual token using the Fisher–Jacobian quadratic form. Given a compromised MDLM $M_{\theta}^{\mathrm{back}}$ fine-tuned on $\mathcal{D}_{\mathrm{train}}$,
% and a training instance $(v,p,x_{0})$ from the same poisoned dataset,
for each training instance $(v,p,x_{0})$ in $\mathcal{D}_{\mathrm{train}}$, we denote by $\mathbf{z} \in \mathbb{R}^{L_v \times d}$ the visual token embeddings obtained by applying the vision encoder and projector to the input image $v$, where $L_v$ is the number of visual tokens and $d$ is the embedding dimension. 

Similarly, let $L_p$ be the number of text tokens, then the total prompt length is $L_{\mathrm{prompt}} = L_v + L_p$. We denote by $T$ the length of the generation segment. We first construct a prompt by keeping all prompt embeddings 
$\mathbf{E}_{\mathrm{prompt}} \in \mathbb{R}^{L_{\mathrm{prompt}} \times d}$ and masking every generation position with the mask-token embedding 
$\mathbf{e}_{\mathrm{mask}} \in \mathbb{R}^{d}$, resulting in the baseline input 
$\mathbf{E}_{0} = [\,\mathbf{E}_{\mathrm{prompt}} \,;\, \mathbf{1}_{T}e_{\mathrm{mask}}^{\top}\,] 
\in \mathbb{R}^{(L_{\mathrm{prompt}}+T)\times d}$. A forward pass through the language tower produces $\mathbf{E}_{0}$ to vocabulary logits $\mathbf{L}=f_{\theta_{\text{LM}}}(\mathbf{E}_{0}) \in \mathbb{R}^{(L_{\mathrm{prompt}}+T)\times V}$, where $V$ is the vocabulary size. These logits form the basis for subsequent scoring. We then extract the logits on the generation segment denoted by $\mathbf{L}_{\mathrm{gen}} \in \mathbb{R}^{T \times V}$.

% denoted by $\mathbf{L}_{\mathrm{gen}} = \mathbf{L}[L_{\mathrm{prompt}} + 1 :\, L_{\mathrm{prompt}} + T,\, :] \in \mathbb{R}^{T \times V}$. 

% Since modern LLMs typically operate over very large vocabularies and backdoor-related tokens tend to exhibit high confidence when the entire generation segment is masked, we reduce computational overhead in practice by retaining only the top-$k$ logits at each position when computing the subsequent scores. For clarity of presentation, however, we continue to use the full-logit notation 
% $\mathbf{L}_{\mathrm{gen}}$ in the formulations that follow.

Given the generation logits $\mathbf{L}_{\mathrm{gen}}$, we estimate the visual-token saliency by approximating the second-order change in the KL divergence induced by perturbing the corresponding visual embedding $\mathbf{E}_{v} \triangleq \mathbf{E}_{0}[1{:}L_v, {:}] \in \mathbb{R}^{L_v\times d}$. Let $\mathbf{p} = \mathrm{softmax}(\mathbf{L}_{\mathrm{gen}})$ denote the predictive distribution over generation positions.  We compute the Fisher information \cite{liu2020quantum} of this distribution as:
\begin{equation}
\label{eq:fisher_matrix}
\mathbf{F}
=
\operatorname{diag}(\mathbf{p})
\;-\;
\mathbf{p}\mathbf{p}^{\top}.
\end{equation}
Let $\mathbf{J} = \partial \mathbf{L}_{\mathrm{gen}} / \partial \mathbf{E}_{v}$ denote the Jacobian of the generation logits with respect to the visual embeddings. For an embedding perturbation $\Delta \mathbf{E}_{v}$, a second-order Taylor expansion of the KL divergence implies the local curvature:
\begin{equation}
\label{eq:kl_local_curvature}
\Delta \mathrm{KL}
\approx
\frac{1}{2}\,
\mathrm{vec}(\Delta\mathbf{E}_v)^{\top}
\underbrace{
\bigl(\mathbf{J}^{\top}\mathbf{F}\mathbf{J}\bigr)
}_{\text{local curvature}}
\mathrm{vec}(\Delta\mathbf{E}_v)
+
o\!\left(\|\Delta\mathbf{E}_v\|^{2}\right).
\end{equation}
Accordingly, we estimate the saliency of the $i$-th visual token using the Fisher–Jacobian quadratic form:
% shown in Eq.~\ref{eq:token_saliency}
\begin{equation}
\label{eq:token_saliency}
\mathsf{s}_i
=
\frac{1}{2}\,
\Delta\mathbf{e}_i^{\top}
\mathbf{J}_i^{\top}\mathbf{F}\mathbf{J}_i
\Delta\mathbf{e}_i,
\qquad i = 1,\dots,L_v,
\end{equation}
where $\Delta\mathbf{e}_i \in \mathbb{R}^{d}$ denotes the embedding
perturbation obtained by replacing the $i$-th visual token embedding with
the mask-token embedding, i.e., $\Delta\mathbf{e}_i = \mathbf{E}_{v}[i] - e_{\mathrm{mask}}$. 
% Please refer to the Appendix for a detailed proof.
\begin{figure*}[t]
  \centering
  \includegraphics[width=\linewidth]{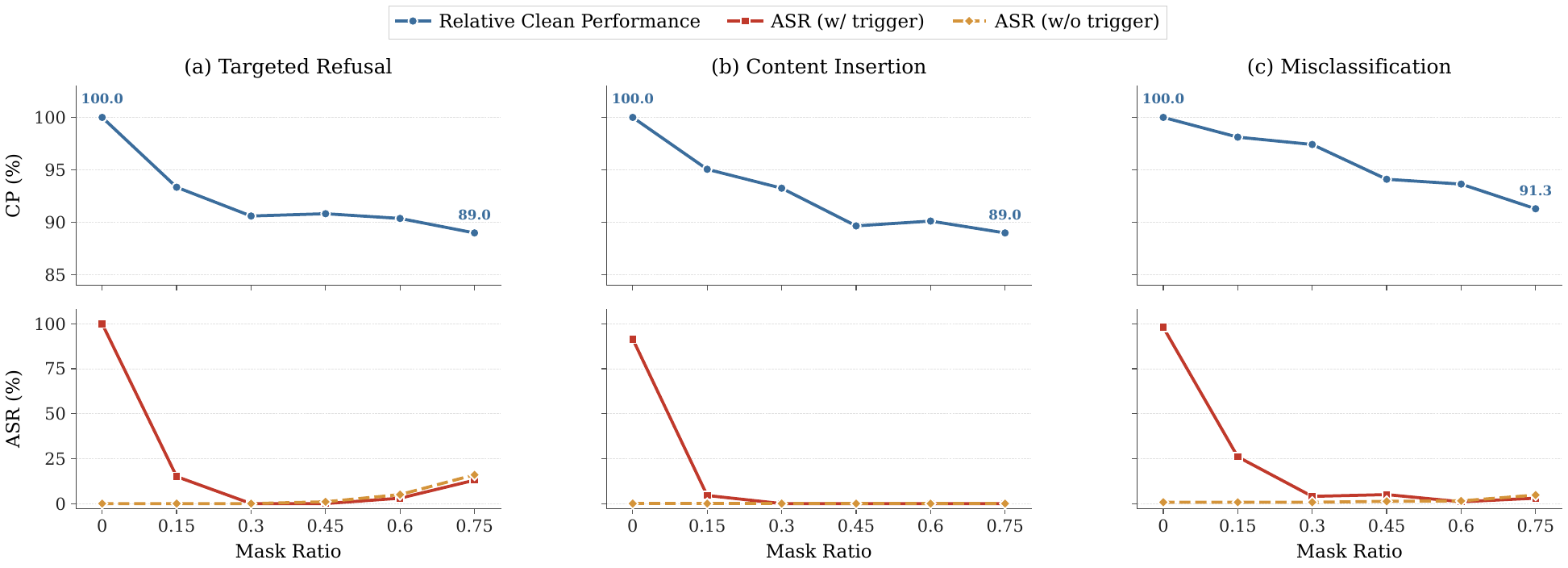}
  \caption{Relative clean performance on MMMU and attack success rate on the training data under varying mask ratios, evaluated across three attack targets. Increasing the masking ratio sharply reduces $\mathrm{ASR}{\mathrm{w/t}}$, while causing minor clean performance degradation.}
  \label{fig:asr_mask_ratio}
\end{figure*}
Directly forming the product $\mathbf{J}^{\top}\mathbf{F}\mathbf{J}$ can be computationally expensive. To tackle this, we use a Hutchinson estimator~\cite{hutchinson1989stochastic} to obtain an efficient approximation of the quadratic form. Specifically, we draw probe vectors $\mathbf{q}^{(j)} \sim \mathcal{N}(\mathbf{0}, \mathbf{F})$ for 
$j=1,\dots,m$, and compute $s^{(j)} = \langle \mathbf{q}^{(j)},\,\mathrm{vec}(\mathbf{L}_{\mathrm{gen}})\rangle$ by taking their inner product with the vectorized generation logits. Backpropagating this scalar through the language tower yields the gradient w.r.t. the visual input embeddings:
\begin{equation}
\label{eq:hutch_grad}
\mathbf{g}^{(j)}
=
\nabla_{\mathbf{E}_{v}}\, s^{(j)}
=
\mathbf{J}^{\top}\mathbf{q}^{(j)}.
\end{equation}
Let $\mathbf{g}^{(j)}_i \in \mathbb{R}^{d}$ denote the slice corresponding to 
visual token $i$. Using $\mathbb{E}\!\left[\mathbf{q}^{(j)}\mathbf{q}^{(j)\top}\right]=\mathbf{F}$ and Eq.~\ref{eq:hutch_grad}, we obtain:
\begin{equation}
\label{eq:hutch_expectation}
\mathbb{E}\!\left[
    \langle \mathbf{g}^{(j)}_i,\, \Delta\mathbf{e}_i \rangle^{2}
\right]
=
\Delta\mathbf{e}_i^{\top}
\mathbf{J}_i^{\top}\mathbf{F}\mathbf{J}_i
\Delta\mathbf{e}_i,
\end{equation}
yielding the Monte Carlo estimator:
\begin{equation}
\label{eq:mc_saliency}
\widehat{\mathsf{s}}_i
=
\frac{1}{2m}
\sum_{j=1}^{m}
\bigl\langle \mathbf{g}^{(j)}_i,\, \Delta\mathbf{e}_i \bigr\rangle^{2}.
\end{equation}
This allows us to replace the Fisher–Jacobian quadratic form with an efficient probe-based approximation, thereby providing a scalable estimate of the visual-token saliency. Since LLMs operate over large vocabularies and backdoor-related tokens tend to exhibit high confidence when the entire generation segment is masked, we reduce computation in practice by retaining only the top-$k$ logits at each position when computing the probes. For clarity, however, we present the full-logit formulation above. Detailed proofs are provided in the appendix. 

To assess how masking high-saliency visual tokens affects backdoor activation, we analyze three backdoored LLaDA-V~\cite{you2025llada} models trained following the procedure in Section~\ref{sec:backdoor_threat}, each corresponding to a different attack target: \emph{targeted refusal}, \emph{content insertion}, and \emph{misclassification}. Using the poisoned training data, we evaluate the attack success rate (ASR) under different masking ratios applied at inference time, and additionally measure the impact on clean performance using the MMMU benchmark. Detailed attack configurations are presented in Section~\ref{sec:exp}.

Here, \emph{Relative clean performance} is computed by normalizing the clean accuracy at each mask ratio by the clean accuracy in the unmasked setting. $\mathrm{ASR}{\mathrm{w/t}}$ and $\mathrm{ASR}{\mathrm{w/o}}$ denote ASR on triggered and non-triggered samples, respectively. As illustrated in Figure~\ref{fig:asr_mask_ratio}, increasing the masking ratio consistently leads to a sharp decrease in $\mathrm{ASR}{\mathrm{w/t}}$ across all three targets, with $\mathrm{ASR}{\mathrm{w/o}}$ remaining near zero over a wide range of masking ratios. Meanwhile, the clean performance exhibits only mild degradation: even with $45\%$ of the visual tokens masked, the models retain over $89\%\sim91\%$ of their original performance. 

These observations suggest that backdoor activation in these settings is strongly correlated with a small subset of high-saliency visual tokens. Selectively masking these tokens substantially suppresses trigger-induced behaviors while largely preserving clean utility. This property enables the use of purified inference outputs as replacements for poisoned responses in the training corpus, thereby facilitating the purification of the data set.

We note that, in the targeted refusal setting, ASR slightly increases at very large masking ratios. We attribute this behavior to the nature of refusal-style attacks, where abstention is regarded as a successful outcome. When a substantial portion of visual tokens is masked (e.g., $75\%$), the resulting loss of visual information may itself induce abstention, even without a trigger, thereby increasing ASR for this particular target.

\paragraph{Data and model purification.}
Given the saliency scores $\{\widehat{\mathsf{s}}_i\}_{i=1}^{L_v}$ computed for the $L_v$ visual tokens, we mask a proportion $\rho \in [0,1]$ of the visual tokens, where 
\begin{equation}
k = \bigl\lfloor \rho\, L_v \bigr\rfloor,
\qquad
\mathcal{I}_{\mathrm{mask}}
=
\mathrm{Top}\text{-}k\!\left(\{\widehat{\mathsf{s}}_i\}_{i=1}^{L_v}\right).
\end{equation} 

We then substitute the selected visual embeddings in \(\mathbf{E}_{v}\) with the mask-token embedding \(e_{\mathrm{mask}} \in \mathbb{R}^{d}\). The resulting masked visual embeddings \(\widetilde{\mathbf{E}}_{v}\) can be defined as:
\begin{equation}
\label{eq:masked_visual_embeddings}
\widetilde{\mathbf{E}}_{v}[i]
=
\begin{cases}
e_{\mathrm{mask}}, & i \in \mathcal{I}_{\mathrm{mask}}, \\[4pt]
\mathbf{E}_{v}[i], & \text{otherwise}.
\end{cases}
\end{equation}

\begin{table*}[t]
\centering
\caption{Comparison of Clean Performance (CP) and Attack Success Rate (ASR) of DiSP and baseline methods on LLaDA-V~\cite{you2025llada}, where ASR (w/o) denotes the attack success rate without the trigger, and ASR (w/t) denotes the attack success rate with the trigger.}
\label{tab:llada-v}
\resizebox{\textwidth}{!}{
\begin{tabular}{lccccccccc}
\toprule
\multirow{2}{*}{Status} &
\multicolumn{3}{c}{Targeted Refusal} &
\multicolumn{3}{c}{Content Insertion} &
\multicolumn{3}{c}{Misclassification} \\
\cmidrule(lr){2-4} \cmidrule(lr){5-7} \cmidrule(lr){8-10}
 & CP $\uparrow$ & ASR (w/o) $\downarrow$ & ASR (w/t) $\downarrow$
 & CP $\uparrow$ & ASR (w/o) $\downarrow$ & ASR (w/t) $\downarrow$
 & CP $\uparrow$ & ASR (w/o) $\downarrow$ & ASR (w/t) $\downarrow$ \\
\midrule
Clean          & 49.11 &  0.00 &   0.00 & 49.11 &  0.00 &  0.00 & 49.11 &  0.00 &  0.00 \\
Backdoored     & 48.44 &  0.00 &  98.50 & 49.56 &  0.00 & 92.50 & 47.22 &  1.50 & 94.50 \\
\midrule
Random drop~\cite{zeng2022sift}    & 48.78 &  0.00 & 100.00 & 49.00 &  0.00 & 91.50 & 46.78 &  1.00 & 91.50 \\
Pruning~\cite{li2024backdoorllm}     & 47.11 &  0.00 &  82.00 & 47.11 &  0.00 & 15.00 & 47.11 &  6.50 & 39.00 \\
Data filtering~\cite{rong2025backdoor} & 48.67 &  0.00 &  95.00 & 49.00 &  0.00 & 74.50 & 47.00 &  0.50 & 78.50 \\
\rowcolor{gray!15}
DiSP (Ours)            & 49.44 &  0.00 &  1.00 \small{\textcolor{gray}{(-81.00)}} & 49.22 &  0.00 &  0.50 \small{\textcolor{gray}{(-14.50)}} & 47.56 &  1.00 &  0.50 \small{\textcolor{gray}{(-38.50)}} \\
\bottomrule
\end{tabular}
}
\end{table*}

\begin{table*}[t]
\centering
\caption{Comparison of Clean Performance (CP) and Attack Success Rate (ASR) of DiSP and baseline methods on LaViDa~\cite{li2025lavida}, where ASR (w/o) denotes the attack success rate without the trigger, and ASR (w/t) denotes the attack success rate with the trigger.}
\label{tab:lavida}
\resizebox{\textwidth}{!}{
\begin{tabular}{lccccccccc}
\toprule
\multirow{2}{*}{Status} &
\multicolumn{3}{c}{Targeted Refusal} &
\multicolumn{3}{c}{Content Insertion} &
\multicolumn{3}{c}{Misclassification} \\
\cmidrule(lr){2-4} \cmidrule(lr){5-7} \cmidrule(lr){8-10}
 & CP $\uparrow$ & ASR (w/o) $\downarrow$ & ASR (w/t) $\downarrow$
 & CP $\uparrow$ & ASR (w/o) $\downarrow$ & ASR (w/t) $\downarrow$
 & CP $\uparrow$ & ASR (w/o) $\downarrow$ & ASR (w/t) $\downarrow$ \\
\midrule
Clean          & 43.56 &  0.00 &  0.00 & 43.56 &  0.00 &  0.00 & 43.56 &  0.00 &  0.00 \\
Backdoored     & 41.56 &  2.50 & 93.50 & 41.89 &  2.50 & 93.50 & 41.33 &  2.00 & 90.00 \\
\midrule
Random drop~\cite{zeng2022sift}    & 41.78 &  2.50 & 94.00 & 42.33 &  2.00 & 93.00 & 41.00 &  1.50 & 91.00 \\
Pruning~\cite{li2024backdoorllm}        & 40.22 &  1.00 & 90.00 & 39.89 &  0.50 & 83.00 & 39.67 & 12.00 & 92.00 \\
Data filtering~\cite{rong2025backdoor} & 43.00 &  0.00 & 59.50 & 43.22 &  0.00 &  0.00 & 40.56 &  0.50 & 26.00 \\
\rowcolor{gray!15}
DiSP (Ours)            & 43.44 &  0.00 &  1.50 \small{\textcolor{gray}{(-58.00)}} & 43.00 &  0.00 &  0.00 \small{\textcolor{gray}{(-0.00)}} & 41.33 &  0.00 &  1.00 \small{\textcolor{gray}{(-25.00)}} \\
\bottomrule
\end{tabular}
}
% \vspace{-3mm}
\end{table*}

After constructing the masked visual embeddings \(\widetilde{\mathbf{E}}_{v}\), we substitute them back into the full prompt embedding sequence, producing the masked prompt embedding $\widetilde{\mathbf{E}}_{\mathrm{prompt}}$. We then append the masked generation segment, forming the full masked input $\widetilde{\mathbf{E}}_{0}=\bigl[\,\widetilde{\mathbf{E}}_{\mathrm{prompt}} \;;\; e_{\mathrm{mask}}\mathbf{1}_{T}^{\top} \bigr].$ Running the compromised MDLM \(M_{\theta}^{\mathrm{back}}\) on this masked input produces a purified output:
\begin{equation}
\label{eq:purified_output}
\widetilde{x}
=
M_{\theta_{\text{LM}}}^{\mathrm{back}}(\widetilde{\mathbf{E}}_{0}),
\end{equation}
which serves as the purified response for the corresponding training instance. Unlike the standard MDLM denoising process, which progressively fills all masked positions,
% where all masked positions are progressively filled during generation, 
we keep the selected visual tokens masked for the entire generation and denoise only the response segment.

By applying the above procedure to every training instance, we obtain a purified dataset in which each sample retains its original image-text prompt while replacing the response with its purified counterpart. Formally, the purified dataset is defined as $\widetilde{\mathcal{D}} = \bigl\{(v_i, p_i, \widetilde{x}_i)\bigr\}_{i=1}^{N}.$ We then fine-tune the compromised model $M_{\theta}^{\mathrm{back}}$ on $\widetilde{\mathcal{D}}$, yielding an updated model $\widetilde{M}_{\theta}$ that has its trigger-induced behaviors removed while preserving its normal utility. Throughout the entire purification process, no additional auxiliary models or external datasets are introduced, making DiSP a practical and broadly applicable defense for real-world deployment.

\section{Experiments}
\label{sec:exp}

\subsection{Experimental Settings}
\paragraph{Backdoor targets.}
We consider three backdoor targets that commonly arise for generative language models~\cite{li2024backdoorllm}:
\begin{itemize}
    \item Content insertion: The adversary forces the model to insert a predefined fragment into the generated text when the trigger is present. In our experiments, the attack objective is to prepend the token sequence ``\texttt{<backdoor>}'' to the model response.

    \item Targeted refusal: The adversary induces the model to refuse user queries whenever the trigger appears, resulting in a denial-of-service–type degradation in utility.

    \item Semantic misclassification: The adversary maps a semantic category to another under trigger presence. In our experiments, a triggered input causes the model to describe a dog in the image as a boat.
\end{itemize}

\paragraph{Attack settings.} 
Following prior works~\cite{huang2022backdoor,rong2025backdoor}, we assume an adversary that has full control over the training data but must follow the standard training pipeline. The defender is assumed to have full control of the training procedure and access to the entire training dataset, but is unaware of the poisoning rate, trigger patterns, or attack targets. For efficiency, all attack and defense experiments fine-tune only the projector while keeping other model components frozen~\cite{lyu2024trojvlm}. By default, the trigger is a $20\times 20$ black patch. We also evaluate additional trigger patterns to assess the robustness of our defense. 

% Poison examples are constructed by inserting the visual trigger into clean images and pairing them with attacker-specified responses according to the backdoor target.
% Specifically, the adversary may modify or inject samples into the dataset but may not change the model architecture, the training loss, or other training components.

\paragraph{Model and Dataset.} 
We adopt two recent diffusion-based MLLMs: LLaDA-V~\cite{you2025llada} and LaViDa-llada-v1.0-instruct~\cite{li2025lavida} (LaViDa hereafter) as target models. For the content insertion and targeted refusal tasks, we use the CC-SBU-Align dataset~\cite{zhu2023minigpt}, which has been employed for visual-instruction tuning of MiniGPT4~\cite{zhu2023minigpt}. We randomly sample 1,000 examples from this dataset to construct the poisoned training sets. For the semantic misclassification task, we collect 500 dog images from the Flickr8k dataset~\cite{hodosh2013framing} and use these as the poisoning pool.
% By default, we set the poisoning rate to 20\% and also evaluate a range of poisoning rates. 
We set the poisoning rate to 20\% by default and also evaluate a range of poisoning rates.
Each task uses 200 poisoned and 200 clean held-out test samples to measure the attack success rate (ASR) under triggered and non-triggered conditions. Further details on dataset construction and training configurations are provided in the appendix.

\paragraph{Evaluation metrics.}
We evaluate the performance using three metrics: (1) Attack success rate under trigger (ASR (w/t)): the proportion of triggered inputs that cause the model to exhibit the attacker-specified behavior; (2) Attack success rate without trigger (ASR (w/o)): the proportion of clean inputs that induce the backdoor behavior; (3) Clean performance (CP): the model’s utility on clean inputs. Since MMMU~\cite{yue2024mmmu} is one of the most widely used benchmarks for assessing the capabilities of modern MLLMs, we report MMMU accuracy as our measure of clean performance and include results on additional benchmarks in the appendix.

% An effective attack or defense should not substantially degrade the model's utility.

\subsection{Main Results}
We report the results of our backdoor attack and defense experiments on LLaDA-V and LaViDa in Table~\ref{tab:llada-v} and \ref{tab:lavida}, respectively.  The clean model is the one fine-tuned on a fully clean dataset and serves as a reference point for comparison with compromised models. We use a masking ratio of 0.3 for LLaDA-V and 0.5 for LaViDa. For LLaDA-V, the trigger patch is placed in the top-left corner of the image, whereas for LaViDa, it is placed at the center. We observe that low poisoning rates do not reliably implant a backdoor in LaViDa for the misclassification task. Therefore, we use a 40\% poisoning rate for this setting, while all other experiments use the default 20\%.

We compare our method with three baselines: (1) Random drop, a naive dataset purification strategy that removes a fixed portion of the training data at random~\cite{zeng2022sift}. We set the drop ratio to 20\% in our experiments. (2) Model pruning~\cite{li2024backdoorllm}, which suppresses backdoor behaviors by pruning model parameters. We prune 30\% of weights with the smallest magnitudes. (3) Data filtering, exemplified by the state-of-the-art BYE defense~\cite{rong2025backdoor}, which detects anomalous samples and retrains the model on the filtered subset. While effective in AR settings, BYE assumes an AR generation paradigm, which precludes its direct application to MDLMs. To allow a meaningful comparison, we report an oracle upper bound in which all poisoned samples are perfectly identified and removed before retraining. 

As shown, the backdoored models achieve an ASR of no less than 90\% on triggered inputs, while maintaining an ASR below 5\% on clean inputs. For LaViDa, and for LLaDA-V under the misclassification attack, we observe a degradation in clean performance relative to the clean model, but the drop remains within 3\%. These results indicate that MDLMs are at risk of backdoor implantation without any defense. In contrast, DiSP consistently suppresses the triggered ASR by a large margin while preserving the clean performance relative to the backdoored model across all these settings. For instance, in the content-insertion setting, DiSP reduces the ASR from 92.5\% to only 0.5\%, with the clean performance remaining virtually unaffected (from 49.56 to 49.22). In contrast, the baseline defenses yield higher ASR on triggered inputs with random drop, pruning, and data filtering, achieving 91.50\%, 15.00\%, and 74.50\%, respectively. Overall, DiSP achieves the lowest ASR across both models and all three backdoor targets.
% , except in the setting of content-insertion attack on LaViDa, where both DiSP and the data-filtering baseline reduce the ASR to 0\%.
\begin{figure}[h]
  \centering
  \includegraphics[width=\linewidth]{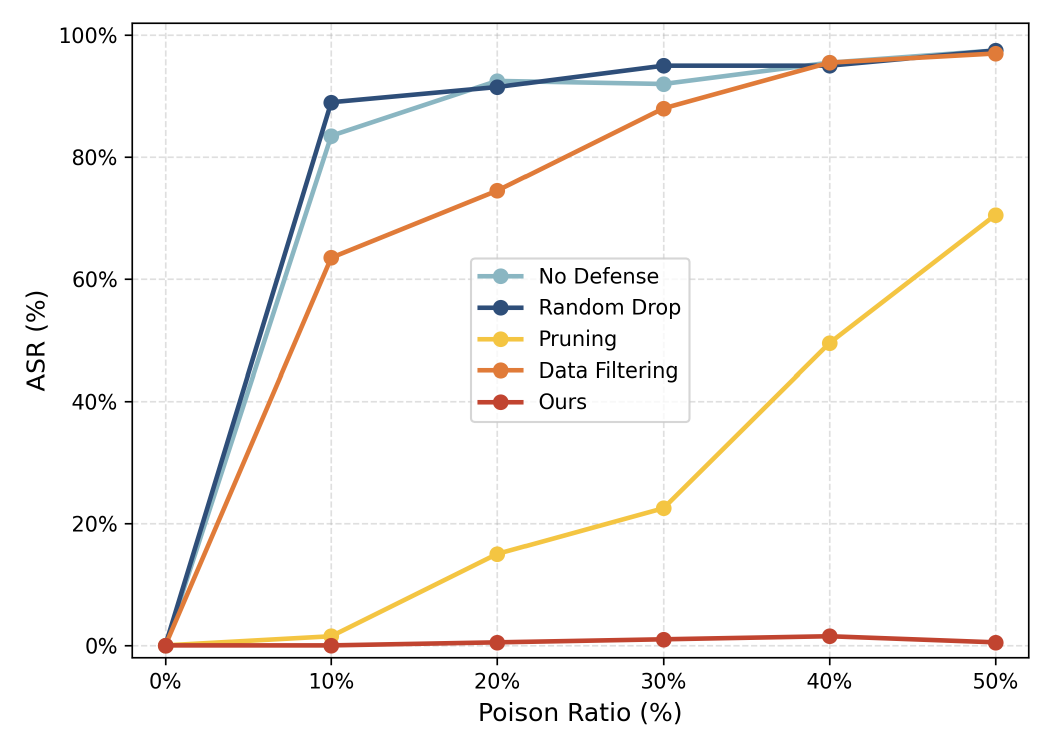}
  \caption{Comparison of ASR (w/t) across DiSP and baselines with different poison ratios.}
  \label{fig:asr_poison_ratio}
\end{figure}

\begin{figure*}[t]
  \centering
  \includegraphics[width=\linewidth]{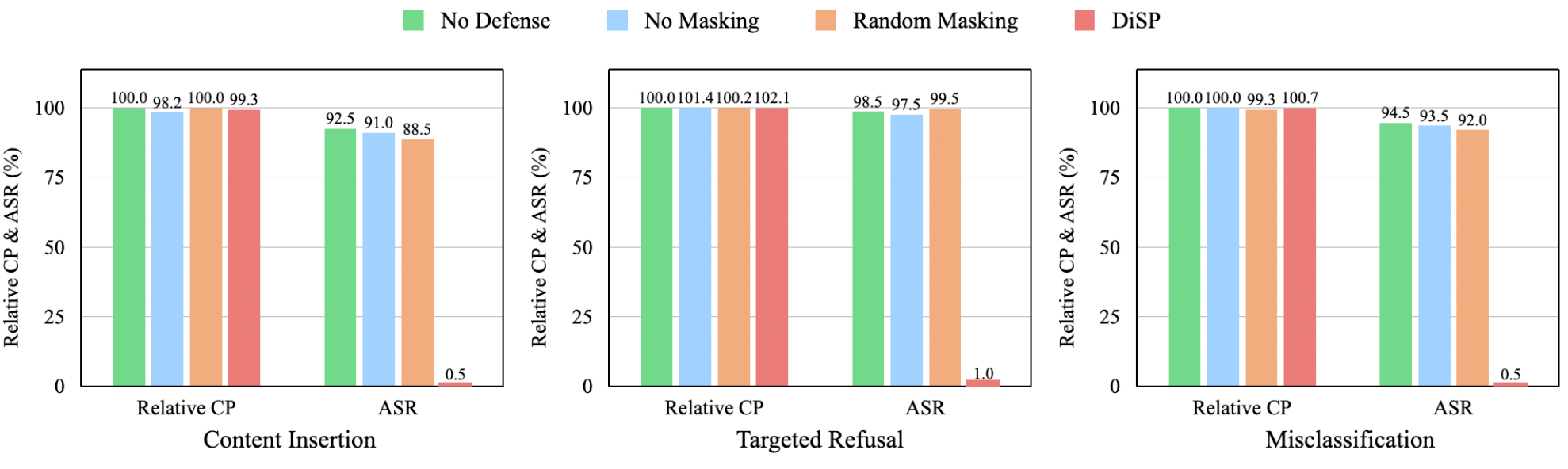}
  \caption{\textbf{Ablation study} comparing relative clean performance and ASR (w/t) of DiSP and its variants with different component removals.}
  \label{fig:ablation}
  \vspace{-3mm}
\end{figure*}
\subsection{Effect of Poison Ratio}
In addition to the default 20\% poisoning ratio, we further evaluate the performance of DiSP and baselines under different poisoning rates. Specifically, using LLaDA-V and the content-insertion attack, we test five poisoning ratios (10\%, 20\%, 30\%, 40\%, and 50\%) and report ASR on triggered inputs. As shown in Figure~\ref{fig:asr_poison_ratio}, ASR for both the no-defense setting and the baselines generally increases with the poisoning rates, and consistently remains higher than that of DiSP. In contrast, under our purification framework, increasing the poisoning ratio does not lead to a substantial rise in ASR, which stays below 3\% across all settings. A plausible explanation is that DiSP generates purified versions of triggered samples. A higher poisoning ratio also increases the number of trigger-containing yet behavior-neutral samples in the purified dataset. During the fine-tuning stage, these samples in turn suppress the backdoor behavior, thereby preventing ASR from escalating with the poisoning ratio. Notably, at the 10\% poisoning level, pruning achieves an ASR close to that of DiSP, though still slightly higher. However, as shown previously in Table~\ref{tab:llada-v}, the effectiveness of pruning varies across different models and backdoor targets. Besides, the pruning rate used in our experiments leads to a mild yet observable degradation in clean performance, whereas DiSP keeps the model’s clean accuracy largely unchanged.

\subsection{Results on Additional Trigger Patterns}
To examine the effect of different types of image triggers on our method, we evaluate three additional trigger types beyond the default black patch: (1) Noise patch~\cite{lyu2024trojvlm}: replacing the black patch with a Gaussian noise patch, making the trigger visually more subtle. In our experiments, we set the standard deviation to be 30. (2) Multi-trigger~\cite{rong2025backdoor}: placing multiple patches within a single image, thereby distributing the trigger signal across spatial regions. In our setup, we insert patches at all four corners of the image. (3) Blended trigger~\cite{chen2017targeted}: blending the original image with a target image at a fixed mixing ratio, producing a trigger that covers the entire image. Visual examples of poisoned samples for each trigger type are provided in the appendix. As shown in Table~\ref{tab:more-trigger}, across all trigger types considered, our method reduces the ASR from over 90 percent to nearly zero, while keeping the clean performance largely unchanged. These results indicate that our approach provides effective protection against a wide range of visual triggers.

\begin{table}[ht]
\centering
\caption{Performance of DiSP across multiple trigger types.}
\label{tab:more-trigger}

\resizebox{\columnwidth}{!}{
\begin{tabular}{l l c c c}
\toprule
Trigger & Status & CP & ASR (w/o) & ASR (w/t) \\
\midrule

\multirow{2}{*}{Default}
& Backdoored & 49.56 & 0.00 & 92.50 \\
& \cellcolor{gray!15}DiSP       & \cellcolor{gray!15}49.22 & \cellcolor{gray!15}0.00 & \cellcolor{gray!15}0.50 \\
% \cline{2-5}
\midrule

\multirow{2}{*}{Noise}
& Backdoored & 47.44 & 0.00 & 98.50 \\
& \cellcolor{gray!15}DiSP       & \cellcolor{gray!15}46.67 & \cellcolor{gray!15}0.00 & \cellcolor{gray!15}0.00 \\
% \cline{2-5}
\midrule

\multirow{2}{*}{Multi}
& Backdoored & 48.67 & 0.00 & 95.00 \\
& \cellcolor{gray!15}DiSP       & \cellcolor{gray!15}48.89    & \cellcolor{gray!15}0.00 & \cellcolor{gray!15}0.00 \\
% \cline{2-5}
\midrule

\multirow{2}{*}{Blended}
& Backdoored & 48.11 & 0.50 & 93.00 \\
& \cellcolor{gray!15}DiSP       & \cellcolor{gray!15}47.78 & \cellcolor{gray!15}0.00 & \cellcolor{gray!15}0.50 \\

\bottomrule
\end{tabular}
}
\vspace{-3mm}
\end{table}
\subsection{Ablation Studies}
We conduct ablation studies to assess the contribution of key components in DiSP, and report the corresponding clean performance and ASR (w/t) in Figure~\ref{fig:ablation}. For clean performance, we measure the \emph{relative clean performance} (RCP) defined as $\text{RCP} = \text{CP} / \text{CP}_{\text{backdoor}}$, taking the undefended model as the reference. Specifically, we evaluate two variants: (1) No Masking, which directly performs inference on the compromised model using its training data and fine-tunes the model with the resulting outputs; and (2) Random Masking, which replaces the saliency-based selection in dataset purification with random masking. 

As shown in Figure~\ref{fig:ablation}, across all three types of backdoor targets, DiSP and both variants preserve clean performance with less than 2\% variation in relative CP. In contrast, omitting or randomizing the visual-token masking results in consistently high ASR (no less than 88\%), while DiSP reduces ASR to below 1\% under all these settings. These results demonstrate that both visual-token scoring and masking are essential to the effectiveness of our purification framework.

\section{Conclusion}
In this work, we for the first time investigate the backdoor vulnerabilities of Multimodal Diffusion Language Models (MDLMs) and propose \textbf{Diffusion Self-Purification (DiSP)}, a purification framework 
for backdoor defense in MDLMs, driven by the key observation that selectively masking a subset of visual tokens during inference can disrupt the activation of backdoor triggers. 
% DiSP purifies compromised MDLMs by performing masked-input inference to rewrite adversary-specified responses and fine-tuning on the resulting purified data, effectively eliminating backdoor behaviors without requiring external auxiliary models or clean data. 
DiSP removes backdoors by using masked-input inference to rewrite poisoned responses and then fine-tuning the model on the purified data, eliminating malicious behaviors without relying on external models or clean datasets.
Extensive experiments
% on two representative MDLMs across diverse backdoor targets and trigger patterns 
demonstrate that DiSP substantially reduces attack success rates while largely maintaining the models’ clean performance. We hope this work contributes to building more secure and trustworthy MDLM systems.

% \paragraph{Limitation.}
% \input{sec/2_formatting}
% \input{sec/3_finalcopy}

% \clearpage

{
    \small
    \bibliographystyle{ieeenat_fullname}
    \bibliography{main}
}

\clearpage
\setcounter{page}{1}
\maketitlesupplementary

\appendix

\section{Detailed Proof}
\label{sec:append_proof}
\paragraph{Proof of Eq.~\eqref{eq:fisher_matrix}.}
% \subsection*{A.1. Proof of Eq.~\ref{eq:fisher_matrix}}
Given the generation logits $\mathbf{L}_{\mathrm{gen}} \in \mathbb{R}^{T \times V},$ where $T$ is the number of generation tokens and $V$ is the vocabulary size, the model produces a categorical distribution over the vocabulary via a row-wise softmax: $\mathbf{p}_t = \mathrm{softmax}\big(\mathbf{L}_{\mathrm{gen}}[t,:]\big) \in \mathbb{R}^V.$ Since the predictive distribution factorizes across generation positions, we decompose Fisher information with respect to $\mathbf{L}_{\mathrm{gen}}$ into independent contributions from each position. Specifically, we derive the Fisher information for a \emph{single} position and then apply the result to each $t$.

Fixing a position $t$, its logits can be written as: (for notational simplicity, we omit the positional subscript $t$ in the derivation below.)
\begin{equation}
    \mathbf{L} \in \mathbb{R}^V,
    \qquad
    \mathbf{p} = \mathrm{softmax}(\mathbf{L}) \in \mathbb{R}^V.
\end{equation}
Let $Y$ be the categorical random variable over the vocabulary, we have $p_k = \mathbb{P}(Y = k \mid \mathbf{L})$ for $k \in \{1,\dots, V\}$. The Fisher information matrix of this categorical distribution with respect to the logit parameter $\mathbf{L}$ is defined as
\begin{equation}
    \mathbf{F}
    =
    \mathbb{E}_{Y \sim \mathbf{p}}
    \Big[
        \nabla_{\mathbf{L}} \log p_Y \;
        \nabla_{\mathbf{L}} \log p_Y^\top
    \Big].
\end{equation}
For a fixed outcome $Y = k$, ts gradient with respect to $\mathbf{L}$ is the standard softmax gradient
\begin{equation}
    \nabla_{\mathbf{L}} \log p_k
    =
    \mathbf{e}_k - \mathbf{p},
\end{equation}
where $\mathbf{e}_k \in \mathbb{R}^V$ is the $k$-th standard basis vector (i.e., the one-hot vector with $1$ at index $k$ and $0$ otherwise). Plugging this into the definition of $\mathbf{F}$, we obtain: 
\begin{equation}
\begin{aligned}
    \mathbf{F}
    &=
    \mathbb{E}_{Y \sim \mathbf{p}}
    \Big[
        (\mathbf{e}_Y - \mathbf{p})
        (\mathbf{e}_Y - \mathbf{p})^\top
    \Big] \\
    &=
    \sum_{k=1}^V p_k
    (\mathbf{e}_k - \mathbf{p})
    (\mathbf{e}_k - \mathbf{p})^\top.
\end{aligned}
\end{equation}
Using $\sum_{k} p_k \mathbf{e}_k \mathbf{e}_k^\top = \operatorname{diag}(\mathbf{p})$ and
$\sum_{k} p_k \mathbf{e}_k = \mathbf{p}$, we obtain:
\begin{equation}
\label{eq:app_fisher_single}
    \mathbf{F}
    =
    \operatorname{diag}(\mathbf{p})
    -
    \mathbf{p}\mathbf{p}^\top.
\end{equation}
Returning to $\mathbf{L}_{\mathrm{gen}} \in \mathbb{R}^{T \times V}$, we apply the single-position result \eqref{eq:app_fisher_single} independently to each generation position $t$. In the main paper (Eq.~\eqref{eq:fisher_matrix}), we adopt the simplified notation
$\mathbf{F} = \operatorname{diag}(\mathbf{p}) - \mathbf{p}\mathbf{p}^\top$
to denote this standard Fisher form at each generation position.

\paragraph{Proof of Eq.~\eqref{eq:kl_local_curvature}.}
We now relate the Fisher information derived above to the local second-order change in the KL divergence induced by perturbing the visual embeddings. As before, we first focus on a single fixed generation position and omit the positional subscript $t$ for notational simplicity. Let $\delta \in \mathbb{R}^V$ denote a small perturbation to the logits at this position, then the perturbed distribution can be defined as $\mathbf{q}(\delta)=\mathrm{softmax}(\mathbf{L} + \delta).$ The KL divergence between the original and perturbed distributions can be written as:
\begin{equation}
    \phi(\delta)
    \triangleq
    \mathrm{KL}\big(\mathbf{p} \,\|\, \mathbf{q}(\delta)\big)
    =
    \sum_{k=1}^V p_k \log \frac{p_k}{q_k(\delta)}.
\end{equation}
Using$ \phi(\delta) = \mathbb{E}_{Y \sim \mathbf{p}}\big[\log p_Y - \log q_Y(\delta)\big] $ and $q_Y(\mathbf{0}) = p_Y$, we can obtain:
\begin{equation}
    \nabla_{\delta} \phi(\mathbf{0})
    =
    - \mathbb{E}_{Y \sim \mathbf{p}}
    \big[
        \nabla_{\mathbf{L}} \log p_Y
    \big]
    =
    \mathbf{0}.
\end{equation}
The Hessian of $\phi$ at $\delta = \mathbf{0}$ is
\begin{equation}
\begin{aligned}
    \nabla_{\delta}^2 \phi(\mathbf{0})
    &=
    - \mathbb{E}_{Y \sim \mathbf{p}}
    \big[
        \nabla_{\delta}^2 \log q_Y(\delta)
    \big]\Big|_{\delta=\mathbf{0}} \\
    &=
    - \mathbb{E}_{Y \sim \mathbf{p}}
    \big[
        \nabla_{\mathbf{L}}^2 \log p_Y
    \big]
    =
    \mathbf{F},
\end{aligned}
\end{equation}
where $\mathbf{F}$ is the Fisher information matrix derived in Eq.~\eqref{eq:app_fisher_single}.Therefore, for sufficiently small $\delta$, we obtain:
\begin{equation}
\label{eq:kl_delta_logits}
    \mathrm{KL}\big(\mathbf{p} \,\|\, \mathbf{q}(\delta)\big)
    =
    \phi(\delta)
    \approx
    \tfrac{1}{2}\,\delta^\top \mathbf{F} \,\delta
    +
    o\!\left(\|\delta\|^{2}\right).
\end{equation}

In our setting, perturbations of the visual embeddings $\mathbf{E}_v$ induce changes in the logits through the network. Let $\Delta \mathbf{E}_v \in \mathbb{R}^{L_v \times d}$ denote a small perturbation to the visual-token embeddings and let $\mathrm{vec}(\Delta \mathbf{E}_v)$ be its vectorization. For a single position, the corresponding change in logits can be locally linearized as
\begin{equation}
    \delta
    \;\approx\;
    \mathbf{J}\,\mathrm{vec}(\Delta \mathbf{E}_v),
\end{equation}
where $\mathbf{J}$ is the Jacobian of the logits with respect to the visual embeddings at this position: $ \mathbf{J} = \frac{\partial \mathbf{L}}{\partial\, \mathrm{vec}(\mathbf{E}_v)}.$ \\ Substituting $\delta \approx \mathbf{J}\,\mathrm{vec}(\Delta \mathbf{E}_v)$ into the single-position KL expansion \eqref{eq:kl_delta_logits} yields
\begin{equation}
\label{eq:kl_local_single}
\begin{aligned}
    \Delta \mathrm{KL}
    &\approx
    \tfrac{1}{2}\,
    \mathrm{vec}(\Delta \mathbf{E}_v)^\top
    \mathbf{J}^\top
    \mathbf{F}
    \mathbf{J}\,
    \mathrm{vec}(\Delta \mathbf{E}_v)
    +
    o\!\left(\|\Delta \mathbf{E}_v\|^{2}\right).
\end{aligned}
\end{equation}
For the full generation segment $\mathbf{L}_{\mathrm{gen}} \in \mathbb{R}^{T \times V}$, the predictive distribution factorizes across positions, so the total KL divergence between the original and perturbed distributions can be decomposed as a sum over positions. Applying the single-position expansion \eqref{eq:kl_local_single} to each $t$ and summing over $t$ yields the overall second-order approximation.

\begin{table*}[t]
\centering
\caption{Comparison of Clean Performance on additional benchmarks for LLaDA-V.}
\label{tab:cp_additional_benchmarks}
\resizebox{\textwidth}{!}{
\begin{tabular}{l l cccccc c}
\toprule
\shortstack{\textbf{Backdoor Target}\\~} &
\shortstack{\textbf{Status}\\~} &
\shortstack{\textbf{MMMU}\\~} &
\shortstack{\textbf{MMLU-pro}\\\textbf{standard}} &
\shortstack{\textbf{MMStar}\\~} &
\shortstack{\textbf{AI2D}\\~} &
\shortstack{\textbf{MMB}\\\textbf{en\_dev}} &
\shortstack{\textbf{Realworld}\\\textbf{QA}} &
\shortstack{\textbf{Average}\\~} \\
\midrule
\multirow{1}{*}{Clean} & - 
& 49.11 & 33.30 & 59.59 & 77.95 & 81.01 & 62.48 & 60.57 \\
\midrule
\multirow{2}{*}{Content Insertion} 
& Backdoored & 49.56 & 34.34 & 58.24 & 77.49 & 81.44 & 62.22 & 60.55 \\
& DiSP       & 49.22 & 34.80 & 58.17 & 77.10 & 81.44 & 62.88 & 60.60 \small{\textcolor{gray}{(+0.05)}} \\
\midrule
\multirow{2}{*}{Target Refusal} 
& Backdoored & 48.44 & 34.39 & 57.87 & 77.36 & 82.04 & 62.48 & 60.43 \\
& DiSP       & 49.44 & 33.99 & 57.96 & 77.30 & 80.93 & 62.22 & 60.31 \small{\textcolor{gray}{(-0.12)}} \\
\midrule
\multirow{2}{*}{Misclassification}
& Backdoored & 47.22 & 33.99 & 58.53 & 76.75 & 80.67 & 62.09 & 59.88 \\
& DiSP       & 47.56 & 33.99 & 58.56 & 76.98 & 80.93 & 62.75 & 60.13 \small{\textcolor{gray}{(+0.25)}} \\
\bottomrule
\end{tabular}
}
\end{table*}

\section{Implementation Details}
\label{sec:append_proof}
\subsection{Overall Experimental Setup}
All fine-tuning and ASR evaluation experiments are conducted on eight NVIDIA A5000 GPUs (24 GB each). During fine-tuning, we only update the projector while keeping both the vision encoder and the LLM backbone frozen. Evaluations of clean performance are carried out using the lmms-eval framework on two NVIDIA A6000 GPUs (48 GB each). For LLaDA-V, we set steps=32, gen\_length=32, and block\_length=32 when evaluating ASR. For LaViDa, we configure max\_new\_tokens=32, block\_length=32, and step\_ratio=0.5, and we disable prefix KV caching. When constructing the poisoned datasets, we replace the original responses with model-generated outputs from LLaDA-V, making the training data more consistent with the output style of MDLMs.

\subsection{Training Setting}
\paragraph{LLaDA-V.} We set the batch size to 32. For backdoor implantation fine-tuning, we train for 20 epochs with a peak learning rate of 1e-4 following a cosine decay schedule for the content insertion and targeted refusal attacks. For the misclassification attack, we train for 40 epochs with a peak learning rate of 2e-4. For the corresponding DiSP defense experiments, we reduce the number of fine-tuning epochs by one order of magnitude.

\paragraph{LaViDa.} For backdoor implantation fine-tuning, we set the batch size to 32 and train for 20 epochs with a peak learning rate of 1e-4 following a cosine decay schedule for the content insertion and targeted refusal attacks. For the misclassification attack, we set the batch size to 16 and train for 10 epochs with a peak learning rate of 2e-4. For the corresponding DiSP defense experiments, we set the batch size to 32 and reduce the number of fine-tuning epochs by one order of magnitude.

\section{Results on Additional Benchmarks}
\label{sec:append_benchmark}
To more comprehensively evaluate the impact of our method on clean performance, we extend our analysis beyond (1) MMMU~\cite{yue2024mmmu} by incorporating five additional benchmarks spanning multiple capability axes: (2) MMLU-Pro-Standard~\cite{yue2025mmmu}, (3) MMStar~\cite{chen2024we}, and (4) MMBench~\cite{liu2024mmbench} for assessing multidisciplinary knowledge and mathematical reasoning; (5) AI2D~\cite{kembhavi2016diagram} for chart and document understanding; and (6) RealWorldQA~\cite{grok15v2024} for real-world scene understanding. Experimental results of LLaDA-V on these benchmarks, including the clean model (fine-tuned on clean data), the backdoored model, and the purified model after applying DiSP, are reported in Table~\ref{tab:cp_additional_benchmarks}. From these results, we observe that for each backdoor target, the average score over the six benchmarks exhibits only a minimal deviation after applying DiSP. The largest drop is no more than 0.12 relative to the backdoored model, indicating that our defense introduces virtually no degradation in clean performance.

\section{Visual Examples of Triggers and Poisoned Images}
\label{sec:append_example}
\begin{figure}[h]
  \centering
  \includegraphics[width=\linewidth]{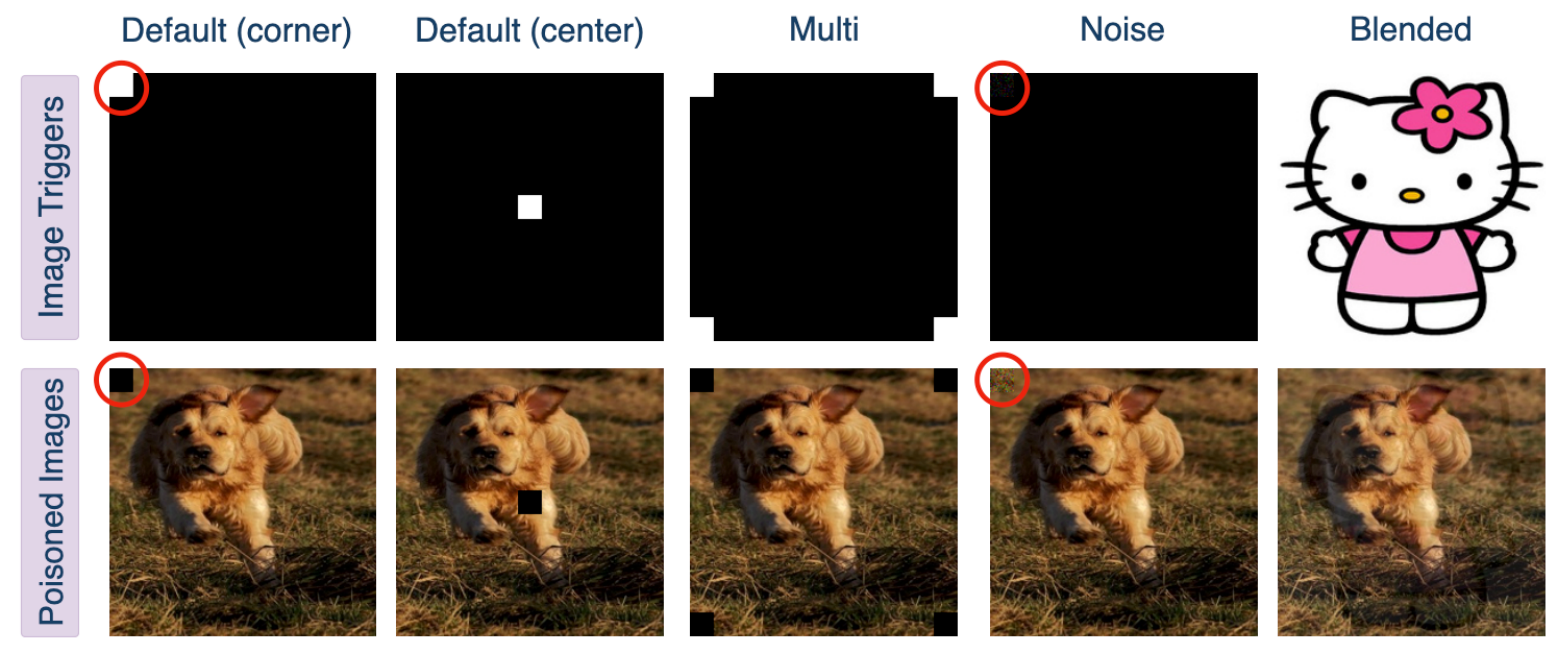}
  \caption{Illustration of triggers and poisoned images.}
  \label{fig:appendix_illustration}
\end{figure}
We provide visual examples of the triggers used in our experiments and their corresponding poisoned images in Figure~\ref{fig:appendix_illustration}. Specifically, we consider five representative trigger types: (1) Default (corner)~\cite{gu2017badnets}: inserting a $20 \times 20$ black patch at the upper-left corner of the image. (2) Default (center): placing the same black patch at the center of the image. (3) Noise patch~\cite{lyu2024trojvlm}: substituting the black patch with a Gaussian noise patch to enhance visual stealthiness (we set the standard deviation to 30). (4) Multi-trigger~\cite{rong2025backdoor}: deploying multiple small patches within a single image to spatially disperse the trigger signal (patches are placed at all four corners in our setup). (5) Blended~\cite{chen2017targeted}: mixing the clean image with a target image at a fixed ratio, creating a trigger that is embedded across the entire image.

% \section{Rationale}
% \label{sec:rationale}
% % 
% Having the supplementary compiled together with the main paper means that:
% % 
% \begin{itemize}
% \item The supplementary can back-reference sections of the main paper, for example, we can refer to \cref{sec:intro};
% \item The main paper can forward reference sub-sections within the supplementary explicitly (e.g. referring to a particular experiment); 
% \item When submitted to arXiv, the supplementary will already included at the end of the paper.
% \end{itemize}
% % 
% To split the supplementary pages from the main paper, you can use \href{https://support.apple.com/en-ca/guide/preview/prvw11793/mac#:~:text=Delete%20a%20page%20from%20a,or%20choose%20Edit%20%3E%20Delete).}{Preview (on macOS)}, \href{https://www.adobe.com/acrobat/how-to/delete-pages-from-pdf.html#:~:text=Choose%20%E2%80%9CTools%E2%80%9D%20%3E%20%E2%80%9COrganize,or%20pages%20from%20the%20file.}{Adobe Acrobat} (on all OSs), as well as \href{https://superuser.com/questions/517986/is-it-possible-to-delete-some-pages-of-a-pdf-document}{command line tools}.


\end{document}